\documentclass[preprint,showpacs,preprintnumbers,amsmath,amssymb]{revtex4}
\usepackage{epsf}
\begin{document}
 
\title{Cold atom realizations of Brownian motors}
  
\author{Ferruccio Renzoni}
  
\affiliation{Departement of Physics and Astronomy, University College London,
Gower Street, London WC1E 6BT, United Kingdom. E-mail: f.renzoni@ucl.ac.uk.
Tel.: 020 7679 7019; Fax. 020 7679 7145}
  
\date{\today}
  
\begin{abstract}
Brownian motors are devices which ``rectify" Brownian motion, i.e. they can
generate a current of particles out of unbiased fluctuations. Brownian
motors are important for the understanding of molecular motors, and are
also promising for the realization of new nanolelectronic devices. Among
the different systems that can be used to study Brownian motors, cold atoms
in optical lattices are quite an unusual one: there is no thermal bath and
both the potential and the fluctuations are determined by laser fields.
In this article recent experimental implementations of Brownian motors using
cold atoms in optical lattices are reviewed. 
\end{abstract}

\maketitle

\section{Introduction}

Brownian motion is the erratic movement of minute particles immersed in a 
fluid \cite{brown,einstein,smolu,langevin}. The erratic motion is generated
by the fluctuating forces acting on the particles, and these forces correspond 
to the collisions between the immersed particles and the fluid constituents. 
Obviously the concept of Brownian motion is not limited to particles immersed
in a fluid, and the description of this phenomenon applies to all those processes 
which can be described as the motion of particles in a fluctuating environment.

Brownian motors \cite{astumian97,bier,astumian02,reimann02,mcclintock02,parrondo} 
are devices which ``rectify" Brownian motion, i.e. they can
generate a current of particles out of unbiased fluctuations. Brownian motors
are attracting much interest from researchers from different areas as they 
seem to have important applications in different fields. In fact, Brownian
motors may represent a model for molecular motors, tiny biological engines
which transform the energy produced in chemical reactions into unidirectional
motion along periodic structures which are macroscopically flat \cite{prost}.
Furthermore, the mechanism of rectification of fluctuations identified in 
the study of Brownian motors may lead to new electron pumps, and indeed the
study of solid state devices which implement Brownian motors is at present a 
very active area of research \cite{linke98,linke,weiss}.

In this article recent experimental implementations of Brownian motors using
cold atoms in optical lattices are reviewed. This is quite an unusual system
for a Brownian motor as there is no a real thermal bath, and both the periodic
potential for the atoms and the fluctuations are determined by laser fields.
With respect to other systems, as for example semiconductor nanostructures,
optical lattices offer the significant advantage of being defect-free and highly
tunable. In fact, the optical potential is determined by the interference of 
laser fields, so it has no defects. And both the potential and the fluctuations
can be carefully controlled by varying the laser fields' parameters.

The present review consists of two parts. The first part introduces the concept 
of the Brownian motor, and discusses the limitations imposed by the second law
of thermodynamics. Then two examples of Brownian motors are presented: the
flashing ratchet and the forced ratchet. The important role of the symmetries
is pointed out. The second part of this work describes the implementation of 
a forced ratchet with cold atoms in an optical lattice. 

\section{Brownian motion}

The concept of Brownian motion can be traced back to 1827, when the Scottish
botanist Robert Brown observed the ceaseless movement of pollen grains
suspended in water. The theoretical description of the phenomenon
came much later, with the work of Einstein in 1905 
\cite{einstein}, Smoluchowski in 1906 \cite{smolu} and Langevin in 1908 
\cite{langevin}.

The Brownian motion of a macroscopically small but microscopically large
particle immersed in a fluid can be simply described in terms of a steady 
dissipative force and a fluctuating force. For a particle of mass $m$ and
position $x(t)$, the equation of motion is
\begin{equation}
m\ddot{x}(t)=-\gamma \dot{x}(t)+f_0\xi (t)~.
\end{equation}
Here $-\gamma \dot{x}(t)$ is the drag force, with $\gamma$ the friction
coefficient; and $f_0\xi (t)$ is a zero-average fluctuating force, which
is generally taken as Gaussian white noise characterised by
\begin{subequations}
\begin{equation}
\langle\xi(t)\rangle = 0 
\end{equation}
\begin{equation}
\langle\xi(t)\xi(t')\rangle = \delta(t-t')~.
\end{equation}
\end{subequations}
The constant $f_0$ and the drag coefficient $\gamma$ are not independent, and
the fluctuation-dissipation relation imposes
\begin{equation}
f_0=\sqrt{2\gamma k_B T}~,
\end{equation}
where $k_B$ is the Boltzmann constant.
Under these assumptions it is possible to show that the mean-square 
displacement $\langle x^2(t)\rangle$ evolves at large times as
\begin{equation}
\langle x^2(t) \rangle = 2 \left( \frac{k_B T}{\gamma} \right) t~.
\end{equation}
This behaviour is termed normal diffusion, with $D=k_B T/\gamma$ the 
diffusion coefficient.

\section{Brownian motors}

\begin{figure}[ht]
\begin{center}
\mbox{\epsfxsize 3.in \epsfbox{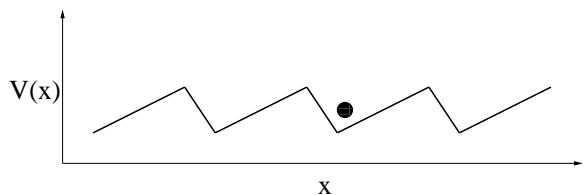}}
\end{center}
\caption{Brownian particle in a ratchet-potential. The potential is periodic,
and has broken spatial symmetry.}
\label{fig1}
\end{figure}

\begin{figure}[ht]
\begin{center}
\mbox{\epsfxsize 1.5in \epsfbox{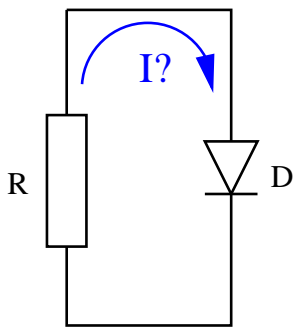}}
\end{center}
\caption{Circuit composed of a resistor and a diode, illustrating the 
Brillouin paradox.}
\label{fig2}
\end{figure}

Brownian motors -- also called ``ratchets" for reasons which will appear clear
in the following -- are devices which rectify fluctuations producing a current.
This is obviously related to the problem of extracting work from fluctuations
and therefore, as pointed out by Feynman \cite{feynman}, the second law of 
thermodynamics imposes strict limitations on the possibility to realize 
Brownian motors. To better illustrate this point, we will refer to two simple
devices, shown in Fig. \ref{fig1} and \ref{fig2}.

Figure \ref{fig1} shows a Brownian particle, in contact with a thermal bath 
(not shown) at temperature $T$, in a periodic asymmetric (``ratchet") potential.
Intuition would suggest directed motion following the rectification of 
fluctuations: the thermal fluctuations induce a random walk of the Brownian 
particle, with the asymmetric potential favouring the motion in one direction, 
acting in this way as a mechanical diode. However this picture is
not correct, and the second law of thermodynamics forbids the generation
of a current.  Indeed the generation of a current through the periodic potential
would imply the possibility of doing work with just one source of
heat at a given temperature, which is forbidden by the second law of
thermodynamics. Therefore, in spite of the broken spatial symmetry no 
preferential direction for the motion arises.  

Another interesting device is shown in Figure \ref{fig2}, and illustrates the
``Brillouin paradox" \cite{brillouin}. This device consists of a resistor connected 
to a diode. The basic idea is very simple. The random thermal motion of the 
electrons in the resistor results in a fluctuating electromotive force across
the resistor, the amplitude of the fluctuations being proportional to the 
temperature $T$ of the resistor and its resistance $R$. In other words, the
resistor $R$ acts as a noise source. The second component of the circuit is a
rectifier (electrical diode), connected in series with the resistor. At first
sight it may seem that the diode would rectify the fluctuating voltage across 
the resistor and produce a directed current through the circuit. However this
is not possible if the resistor and the diode are at the same temperature.
As for the case examined previously, the generation of a directed current through
the circuit would imply the possibility of doing work with just one source of 
heat at a given temperature, which is forbidden by the second law of
thermodynamics. 

At this point one may wonder whether it is possible at all to realize a Brownian
motor, i.e. a device which rectifies fluctuations producing a direct current. 
The answer is affirmative, provided that the second law of thermodynamics
is not violated, i.e. provided that the rectification of fluctuations does
not correspond to the extraction of work from just one source of heat at one 
temperature. As we will see, this can be realized by driving a system out of 
equilibrium. We will discuss now two examples of Brownian motors realized
in this way: the flashing ratchet and the forced ratchet.

\subsection{The flashing ratchet}

\begin{figure}[ht]
\begin{center}
\mbox{\epsfxsize 4.5in \epsfbox{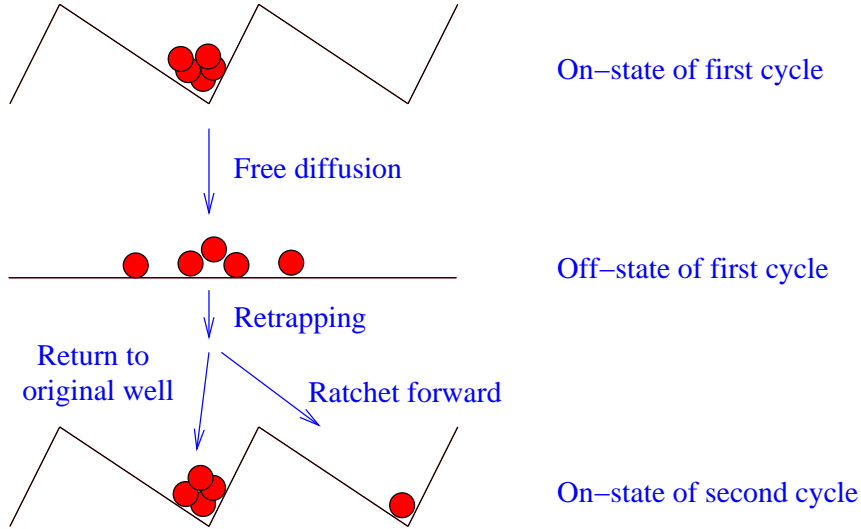}}
\end{center}
\caption{Working principle of the flashing ratchet.}
\label{fig3}
\end{figure}

The working principle of the flashing ratchet \cite{adjari} is illustrated in
Fig.~\ref{fig3}. A sample of Brownian particles experiences an asymmetric
potential, analogous to the system described in the previous Section and 
illustrated in Fig.~\ref{fig1}. The difference between the present 
system and the device of Fig.~\ref{fig1} is that now the potential is 
intermittent. Indeed the potential is turned on and off repeatedly, either
periodically or at random instants. The working prionciple is quite simple
(see Fig.~\ref{fig3}).

Consider an initial situation with the potential turned on and the Brownian 
particles localized at the bottom of a given well. Then the potential is 
turned off, and the Brownian particles will symmetrically diffuse in space. 
Then the potential is turned on again, and the Brownian particles are retrapped
in both the original well and in a few neighbouring ones. However, as the 
potential is asymmetric the retrapping will lead to an asymmetric situation, 
with the number of particles trapped in the wells at the left of the original
well different from the number of particles trapped in the wells  at the right
of the starting location. Indeed it is clear from Fig.~\ref{fig3} that the 
wells closer to the ``steep wall" of the starting well will collect more particles 
during the retrapping phase. In this way the center of mass of the particle 
cloud will move, and directed motion is thus obtained.

It is important to point out why the operation of the flashing ratchet does 
not violate the second law of thermodynamics. This is because work is done
on the system while turning on the potential. Thus, although fluctuations 
are rectified and a current is generated, this does not imply that work has 
been extracted out of just one heat source as some additional work was
necessary to turn on the potential. Therefore the second law of 
thermodynamics is not violated.

A final remark on the efficiency of the flashing ratchet. A detailed analysis 
of the dependence of the current amplitude on the flashing ratchet parameters
is beyond the scope of the present article, and we refer the interested reader
to Refs.~\cite{astumian97,bier}.

\subsection{The forced ratchet}

\begin{figure}[ht]
\begin{center}
\mbox{\epsfxsize 2.in \epsfbox{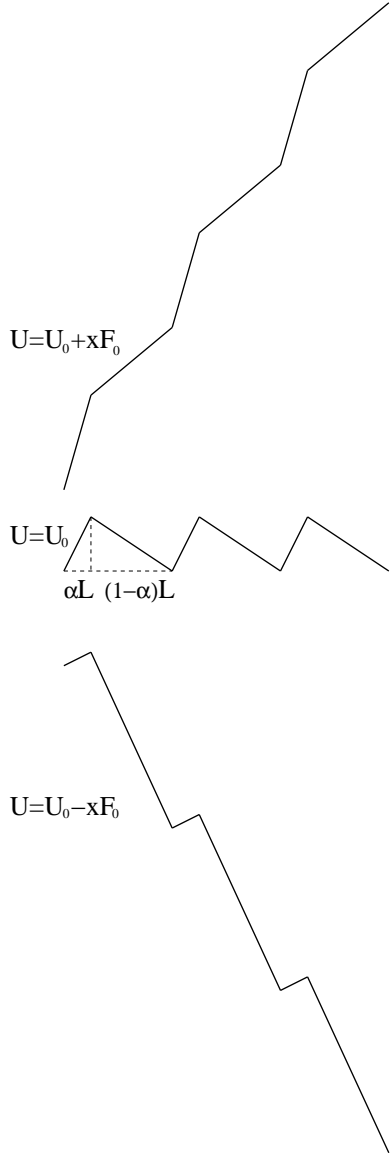}}
\end{center}
\caption{Forced ratchet potential.}
\label{fig4}
\end{figure}

A second example of Brownian motor is the forced (or rocking) ratchet 
(Fig.~\ref{fig4}) \cite{magnasco,bartussek}.

Consider once again a sample of Brownian particles in an asymmetric potential,
$U_0$, with parameters as in Fig.~\ref{fig4}. The potential is now rocked by
an applied force $F$ fluctuating between $-F_0$ and $+F_0$ with, say,
$F_0 > 0$. The relevant parameters to which compare $F_0$ are the two slopes
of the piecewise potential $U_0$: $\Delta U/[(1-\alpha )L]$ and 
$\Delta U/[\alpha L]$. Two different cases are relevant for the present 
analysis. First we examine the case for 
$\Delta U/[(1-\alpha )L]< F_0 < \Delta U/[\alpha L]$. This is the situation
represented in Fig.~\ref{fig4}. When the applied force equals $-F_0$ the 
potential $U=U_0+xF_0$ increases monotonically for increasing $x$, while
when the applied force equals $+F_0$ the total potential $U=U_0-xF_0$ 
still shows potential minima where the particles can be trapped. It appears
evident that a force slowly oscillating between $-F_0$ and $+F_0$ will lead 
to directed motion as for $F=-F_0$ the potential is monotonically increasing
while for $F=+F_0$ the potential shows wells which allow the trapping of 
particles. A current in the negative direction is thus produced. We note 
however that this current is produced also in the absence of fluctuations,
and therefore the current cannot be attributed entirely to the rectification
of fluctuations.

A more relevant case for the Brownian motors is the subthreshold regime
$F_0<\Delta U/[(1-\alpha )L]$. In this case the rocking potential alone
is not sufficient to induce directed motion. However the presence of some
thermal noise allows the generation of a current of particles, and the device
acts therefore as a fluctuations rectifier. In this case the current is 
entirely due to the rectification of fluctuations, and the rocking ratchet is a realization of a Brownian motor.

\subsection{The rocking ratchet: role of the symmetries}\label{rock}

In the previous Sections we have seen two examples of Brownian motors, both
based on an asymmetric (``ratchet") potential. We should mention here that 
the term  "ratchet" is often used as synonymous of Brownian motor, or
fluctuations rectifier, even if, as we will see in this Section, Brownian 
motors can also be realized by using spatially symmetric potentials
\cite{goychuk,flach00,flach01,flach02,dykman}.

In this Section we identify the conditions to observe directed motion in a 
rocking ratchet, pointing out the role of the symmetries. This will constitute
the basis for the cold atom realization of a Brownian motor described in the
following Sections.

For the sake of simplicity, we limit our analysis of the role of the symmetries
to the case of weak damping of the particles. We refer the interested reader
to Refs. \cite{super,flach01,flach02} for a discussion of the modification of the
symmetries of a ratchet-device with increasing dissipation.

Omitting completely the dissipation term, the equation of motion for the
particle of mass $m$ can be written as:
\begin{equation}
m\ddot{x}(t)=-U'(x)+E(t)~.
\label{eq_motion}
\end{equation}
Here $U$ is a spatially periodic potential of period $\lambda$: 
$U(x)=U(x+\lambda)$, and $E$ a zero mean ac field of period $T$: 
$E(t)=E(t+T)$.

Now, we are interested in determining the symmetries which forbid the
appearance of directed motion. These correspond to the transformations
in $x,t$ which produce a change in sign of the momentum $p$, i.e. 
the transformations which map a trajectory $\{ x(t;x_0,p_0), p(t;x_0,p_0)\}$,
with $x_0, p_0$ the initial position and momentum, into another one with
opposite momentum. These transformations consist of reflections and shifts 
in time and space and are \cite{flach00,super,flach01,flach02}:

\begin{subequations}
\begin{equation}
\hat{S}_a  
\begin{pmatrix} x(t;x_0,p_0) \\ p(t;x_0,p_0) \end{pmatrix} =
\begin{pmatrix} -x(t+T/2;x_0,p_0)+2\chi \\ -p(t+T/2;x_0,p_0) \end{pmatrix} 
\end{equation}
\begin{equation}
\hat{S}_b 
\begin{pmatrix} x(t;x_0,p_0) \\ p(t;x_0,p_0) \end{pmatrix} =
\begin{pmatrix} x(-t+2\tau;x_0,p_0) \\ -p(-t+2\tau;x_0,p_0) \end{pmatrix} 
\end{equation}
\end{subequations}
with $\chi$ and $\tau$ constants. Clearly, if the equation of motion Eq.~\ref{eq_motion}
is invariant under the transformations $\hat{S}_a$,  $\hat{S}_b$ then no current can
be generated. Whether $\hat{S}_a$,  $\hat{S}_b$ are symmetries of the system depends 
on the form of $U(x)$ and $E(t)$. It can be shown that \cite{flach00,super,flach01,flach02}:

\begin{description}
\item[A)] 
If $U'(x+\chi)=-U'(-x+\chi)$ and $E(t)=-E(t+T/2)$ then the equation of motion 
Eq.~\ref{eq_motion} is invariant under $\hat{S}_a$, and no directed motion can be observed;
\item[B)] 
If $E(t+\tau)=E(-t+\tau)$ the system is invariant under the transformation $\hat{S}_b$ and 
no current can be generated.
\end{description}

In conclusion, the possibility to generate a current is closely related to the symmetries 
of the system. In the remaining of this paper we will present explicit examples of how the
breaking of the relevant symmetries leads to the generation of a current.

\subsection{The spatially symmetric rocking ratchet}\label{symmetric}

In the following we will present in detail a cold atom realization of a spatially symmetric
rocking ratchet. It is therefore interesting to analyze in the present context the role of
the symmetries in that specific case.

We consider here a spatially periodic potential
\begin{equation}
U(x)=-\cos x
\end{equation}
and a bi-harmonic periodic driving
\begin{equation}
E(t) = E_1 \cos (\omega t) + E_2 \cos (2\omega t + \phi) ~.
\end{equation}

If $E_1\neq 0$ and $E_2\neq 0$ the symmetry $\hat{S}_a$ is broken as $E(t)\neq -E(t+T/2)$,
independently of the value of $\phi$. On the other hand, always under the assumption 
that $E_1\neq 0$ and $E_2\neq 0$, the symmetry of the system under the transformation
$\hat{S}_b$ is controlled by the relative phase $\phi$: $\hat{S}_b$ is a symmetry of
the system if $\phi=n\pi$, with $n$ integer. To summarize, in the presence of both
harmonics the symmetry of the system is controlled by the relative phase $\phi$, and indeed
it can be shown \cite{flach01} that a current appears, whose magnitude $C$ is proportional to
\begin{equation}
C\propto \frac{E_1^2 E_2}{\omega^3}\sin\phi ~.
\end{equation}
We notice that the current vanishes when $\phi=n\pi$, as the symmetry $\hat{S}_b$ is restored.

We stress that the system considered here is purely Hamiltonian, and neither damping nor
fluctuations were included in the model. The current is the result of non-linear 
mixing of the two harmonics, and does not correspond to the realization of a Brownian 
motor. Theoretical work already examined this system in the presence of dissipation
\cite{flach01}. In the present work we do not repeat the theoretical derivation, and 
follow a less formal approach: the role of dissipation will be discussed in next
Section, where experimental realizations of the rocked ratchet will be reported and
it will be shown how the addition of fluctuating forces modifies the picture discussed 
in the present Section and leads to the realization of a Brownian motor.

\section{Cold atom realization of a Brownian motor}

The first experiment on the ratchet effect using cold atoms in an optical lattice
was reported by Mennerat-Robilliard {\it et al} \cite{cecile}. In that work directed motion
was observed in a spatially asymmetric dark optical lattice. In the present paper
we describe in detail a different cold atom realization of a Brownian motor, with
a forced ratchet realized by using atoms in a bright optical lattice. This is a
direct implementation of the ideas discussed in Secs. \ref{rock} and \ref{symmetric}. 
Before entering into the details of the realization of the forced ratchet, we summarize 
the basics of bright optical lattices and the underlying Sisyphus cooling mechanism.
For more comprehensive reviews of optical lattices, we refer the reader to Refs.
\cite{meacher,verkerk,robi}.

\subsection{Bright optical lattices}\label{bright}

Optical lattices are periodic potentials for atoms created by the interference
of two or more laser fields. In near-resonant optical lattices a set of laser
fields produce at once the periodic potential for the atoms and the cooling
mechanism, named Sisyphus cooling, which decreases their kinetic energy so
that they are finally trapped at the bottom of the potential wells. We describe
here the principles of these optical lattices in the case of a one-dimensional
configuration and a $J_g=1/2\to J_e=3/2$ atomic transition. This
is the simplest configuration in which Sisyphus cooling takes place.

\begin{figure}[ht]
\begin{center}
\mbox{\epsfxsize 3.in \epsfbox{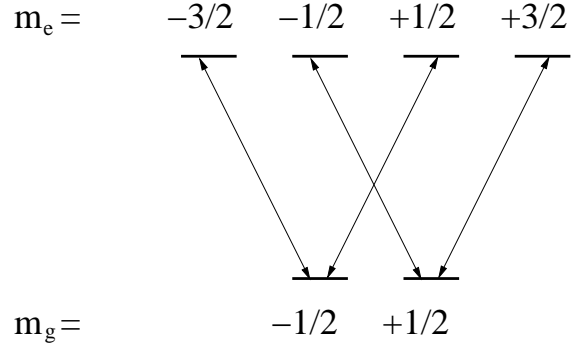}}
\end{center}
\caption{Atomic level scheme for a $J_g=1/2\to J_e=3/2$ transition. The arrows
indicate the couplings due to $\sigma^{+}$, $\sigma^{-}$ laser excitation.}
\label{fig5}
\end{figure}

\begin{figure}[ht]
\begin{center}
\mbox{\epsfxsize 3.in \epsfbox{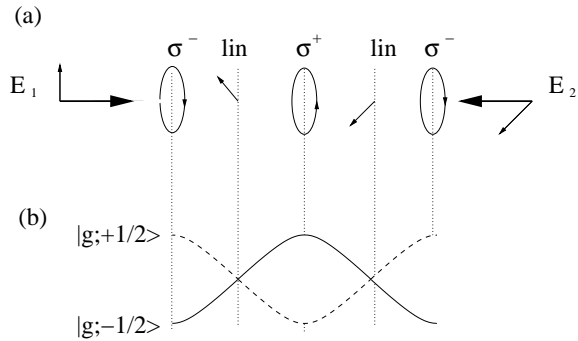}}
\end{center}
\caption{(a) Arrangement of laser fields in the so-called lin$\perp$lin 
configuration, and resulting gradient of ellipticity. (b) Light shift of the
two ground-state Zeeman sublevels $|g,\pm 1/2\rangle$.}
\label{fig6}
\end{figure}

Consider a transition $J_g=1/2\to J_e=3/2$ (Fig.~\ref{fig5}) coupled to two
laser fields with the same amplitude and the same wavelength $\lambda$, linearly
polarized and counterpropagating. These laser fields are detuned below atomic 
resonance and have orthogonal linear polarization (lin$\perp$lin configuration,
see Fig.~\ref{fig6}(a)):
\begin{subequations}
\begin{equation}
\vec{E}_1(z,t) = \frac{1}{2}\vec{\epsilon}_x E_0 \exp [i(kz-\omega t)]+c.c 
\end{equation}
\begin{equation}
\vec{E}_2(z,t) = \frac{1}{2}\vec{\epsilon}_y E_0 \exp [i(-kz-\omega t+\alpha)]+c.c 
\end{equation}
\end{subequations}
where $k=2\pi/\lambda$ and $\omega=kc$ are the laser field wavevector and
angular frequency, respectively. The total electric field is 
\begin{equation}
\vec{E}_1(z,t)+\vec{E}_2(z,t) = [E_{+}(z)\vec{\epsilon}_{+}+
E_{-}(z)\vec{\epsilon}_{-}]\exp (-i\omega t)+c.c.
\end{equation}
where $\vec{\epsilon}_{\pm}$ are the unit vectors of circular polarization
and, after elimination of the relative phase $\alpha$ through an appropriate
choice of the origin of the space- and time-coordinates, $E_{+}$ and 
$E_{-}$ are given by: 
\begin{subequations}
\label{eq_pol}
\begin{equation}
E_{+} = -i\frac{E_0}{\sqrt{2}} \sin kz ~,
\end{equation}
\begin{equation}
E_{-} = \frac{E_0}{\sqrt{2}} \cos kz~.
\end{equation}
\end{subequations}
The superposition of the two laser fields $E_1$, $E_2$ produces therefore an 
electric field characterized by a constant intensity and a spatial gradient of 
polarization ellipticity of period $\lambda/2$, as shown in Fig.~\ref{fig6}(a).

We examine now the effects that the laser fields have on the atoms. The basic
mechanism responsible for the generation of a periodic potential is the 
"light shift": a laser field coupling a given transition, and characterized
by an intensity $I_L$ and detuning $\Delta$ from atomic resonance, leads to 
a shift of the ground state energy ("light shift") proportional to $I/\Delta$.

In the present case of a $J_g=1/2\to J_e=3/2$ transition there are two laser fields 
coupling each ground state sublevel to the excited state, and contributions from all these 
couplings have to be taken into account to derive the light shifts $U_{\pm}$ for 
the ground state Zeeman sublevels $|g,\pm 1/2\rangle$. We will omit here the details
of the calculations, and simply report the final results for the light shifts 
(see Ref.~\cite{robi} for the derivation):
\begin{subequations}
\begin{equation}
U_{+} = 2\hbar\Delta_0' \left( \frac{I^{+}}{I} + \frac{I^{-}}{3I} \right)~,
\end{equation}
\begin{equation}
U_{-} = 2\hbar\Delta_0' \left( \frac{I^{-}}{I} + \frac{I^{+}}{3I} \right)~.
\end{equation}
\end{subequations}
Here $I^{\pm}=|E^{\pm}|^2$ are the intensities of the right- and left-polarization 
components of the light, and $I=I^{-}+I^{+}$ is the total intensity. The quantity 
$\Delta_0'$ is the {\it light shift per beam}, equal to
\begin{equation}
\Delta_0'=\Delta\frac{\Omega_1^2/4}{\Delta^2+\Gamma^2/4},
\end{equation}
where $\Delta$ is the detuning of the optical field from atomic resonance, $\Gamma$ the
linewidth of the atomic transition and $\Omega_1$ the resonant Rabi frequency for a 
transition having a Clebsch-Gordan coefficient equal to 1. The square of the resonant
Rabi frequency is proportional to the light intensity, so in the limit of not too small
detuning $\Delta$, we find that the light shift per beam scales as $I/\Delta$, as already
mentioned. By substituting the expressions (\ref{eq_pol}) for $E^{+}$, $E^{-}$, the light
shifts $U_{\pm}$ can be rewritten as:
\begin{eqnarray}
U_{\pm} = \frac{U_0}{2} [ -2 \pm \cos kz ]
\end{eqnarray}
with
\begin{equation}
U_0 = -\frac{4}{3}\hbar \Delta_0'
\end{equation}
the depth of the potential wells. We therefore conclude that the light ellipticity 
gradient produces a periodic modulation of the light shifts of the ground state
Zeeman sublevels (Fig.~\ref{fig6}(b)). These periodic modulation acts as an optical 
potential for the atoms, and indeed these periodically modulated light shifts are
usually referred to as {\it optical potentials}.

We turn now to the analysis of the cooling mechanism, the so-called Sisyphus 
cooling \cite{sisifo}, which decreases the kinetic energy of the atoms and allows 
their trapping at the bottom of the wells of the optical potential.

\begin{figure}[ht]
\begin{center}
\mbox{\epsfxsize 3.in \epsfbox{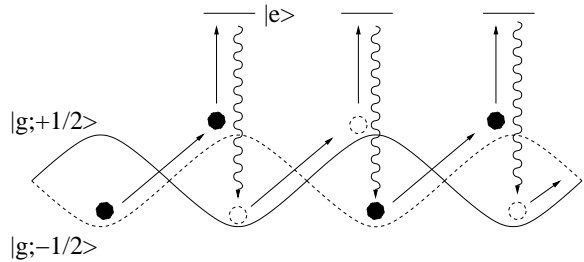}}
\end{center}
\caption{Sisyphus cooling mechanism.}
\label{fig7}
\end{figure}

Sisyphus cooling is determined by the combined action of the light shifts and of
optical pumping, which transfers, through cycles of absorption/spontaneous emission,
atoms from one ground state sublevel to the other one. This is illustrated in
Fig.~\ref{fig7}. Consider an atom moving 
with a positive velocity, and initially at $z=0$ in the state $|g,-1/2\rangle$.
While moving in the positive $z$ direction the atom climbs the potential curve
corresponding to its actual internal state. This has two consequences: first, a 
part of the kinetic energy of the atoms is transformed in potential energy; 
second, the component $\sigma^{+}$ of the light increases, which implies the 
increase of the optical pumping rate towards the level $|g,+1/2\rangle$, i.e.
an increase of the probability of transfering the atom from the actual internal
state $|g,-1/2\rangle$ to the state $|g,+1/2\rangle$. At the top of the 
potential hill ($z=\lambda/4$, see Fig.~\ref{fig6}) the polarization of the 
light is purely $\sigma^{+}$, and the probability to transfer the atom into the 
sublevel $|g,+1/2\rangle$ is very large. The transfer of the atom into the 
level $|g,+1/2\rangle$ results into a loss of potential energy, which is 
carried away by the spontaneously emitted photon. This process is repeated 
several times, until the atom does not have enough energy any more to reach the
top of a potential hill, and it is trapped in a well. We notice here the
analogy with the myth of Sisyphus, king of Corinth, condemned forever to roll
a huge stone up a hill which repeatedly rolls back to the bottom before the 
summit is reached. This is why the described cooling mechanism has been named
Sisyphus cooling. The described cooling process leads to the localization of the atoms at 
the bottom of the potential wells, and we obtain in this way an {\it optical 
lattice}: an ensemble of atoms localized in a periodic potential. We notice 
that the atoms are localized at the sites where their interaction with the light 
is maximum. It is because of this property that optical lattices of this type are 
termed {\it bright} optical lattices.

\subsection{Realization of the forced ratchet}

\begin{figure}[ht]
\begin{center}
\mbox{\epsfxsize 2.5in \epsfbox{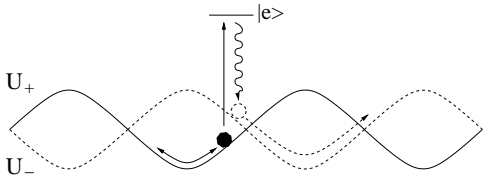}}
\end{center}
\caption{Stochastic process of optical pumping transferring, via an excited 
state, an atom from a potential to the other one. The filled (empty) circle
represents the atom in the $|g,+1/2\rangle$ ($|g,-1/2\rangle$) ground state
sublevel.}
\label{fig8}
\end{figure}

The realization of a forced ratchet requires essentially three elements. First, 
a periodic potential; second, a fluctuating environment which results in friction 
and in a fluctuating force. Finally, it should be possible to apply a zero-mean 
ac-force to the particles (the atoms in the present case). All these requirements
can be satisfied by using cold atoms in optical lattices, as it was demonstrated
in Ref.~\cite{schiavoni}. In that work the one-dimensional lin$\perp$lin optical
lattice described in Sec.~\ref{bright} was taken as periodic potential. We stress
that such an optical lattice is spatially symmetric, and indeed the work of 
Ref.~\cite{schiavoni} aimed to realize the spatially symmetric rocking ratchet 
introduced in Sec.~\ref{rock}. We turn now to the analysis of the friction and 
fluctuations in the optical lattice, the second element necessary to use optical
lattices as a model system for Brownian motors. As already discussed, the optical 
pumping between the different atomic ground state sublevels combined with the spatial 
modulation of the optical potential leads to the cooling of the atoms and to 
their localization at the minima of the optical potential. The essential fact 
for the realization of Brownian motors is that even after the cooling phase, 
charaterized by a decrease of the kinetic energy of the atoms and their trapping 
in the optical potential, the atoms keep interacting with the light fields and 
this induces fluctuations in the atomic dynamics. Indeed, consider an atom that
has already lost enough energy to be trapped at the bottom of a potential well. 
The atom will then oscillate at the bottom of the well. This is the situation 
illustrated  in Fig.~\ref{fig8}.

To be specific, consider
for example an atom initially in the $|g,+1/2\rangle$  state. Exactly at the 
center of the well the light polarization is purely $\sigma^{+}$, which does 
not allow the transfer from the $|g,+1/2\rangle$ state to the $|g,-1/2\rangle$ 
sublevel. However, out of the center of the well the light has also a nonzero
$\sigma^{-}$ component, which results in a nonzero probability to transfer 
the atom from its original sublevel to the other one. Therefore the atom 
can be transfered from one sublevel to the other one, and also the potential 
experienced by the atom will change from $U_{+}$ to $U_{-}$, i.e. the force
experienced by the atom will change. As optical pumping is a stochastic process,
the (stochastic) transfer from a sublevel to the other one results in a 
{\it fluctuating force}. Figure \ref{fig8} also shows how optical 
pumping between different optical potentials leads to the transport of atoms
through the lattice: although the trapped atom does not have enough energy to 
climb the potential hill, optical pumping allows the transfer from a potential 
well to the neighbouring one. The optical pumping leads then to a {\it random walk} 
of the atoms through the optical potential, and indeed normal diffusion has 
been experimentally observed for an atomic cloud expanding in an optical
lattice \cite{regis}.  

The last element necessary to implement a rocking ratchet is the oscillating
force. In order to generate a time-dependent homogeneous force, one of the 
lattice beams is phase modulated, so that to obtain the electric field 
configuration:
\begin{equation}
\frac{1}{2} E_0 \left\{ \vec{\epsilon}_x \exp [i(kz-\omega t)] +
\vec{\epsilon}_y E_0 \exp [i(-kz-\omega t+\alpha (t) )]\right\} +c.c.~,
\end{equation}
where $\alpha (t)$ is the time-dependent phase. In the laboratory reference
frame this laser configuration generates a moving optical potential 
$U[2kz-\alpha(t)]$. Consider now the dynamics in the moving reference frame 
defined by $z'=z-\alpha(t)/2k$. In this accelerated reference frame the optical 
potential is stationary. In addition to the potential the atom, of mass $m$, 
experiences also an inertial force $F$ in the $z$ direction proportional to 
the acceleration $a$ of the moving frame \cite{landau}:
\begin{equation}
F=-ma=\frac{m}{2k}\ddot{\alpha}(t)~.
\label{inertial}
\end{equation}
In this way in the accelerated frame of the optical potential the atoms 
experience an homogeneous force which can be controlled by varying the phase
$\alpha (t)$ of one of the lattice beams. The appropriate choice of the 
phase $\alpha(t)$ for the realization of the spatially symmetric rocking 
ratchet is
\begin{equation}
\alpha(t) = \alpha_0 \left[ A\cos (\omega t) +\frac{B}{4} \cos (2\omega t-\phi)
\right]
\label{alpha}
\end{equation} 
with $\phi$ constant. Indeed, by using Eq.~\ref{inertial}, we can see immediately
that in the accelerated frame of the optical potential the phase modulation 
$\alpha(t)$ will result into a force
\begin{equation}
F=\frac{m\omega^2\alpha_0}{2k}\left[ A\cos (\omega t) + B \cos (2\omega t-\phi)
\right]
\end{equation}
which is of the form needed for the realization of the spatially symmetric 
rocking ratchet, as discussed in Sec. \ref{symmetric}.

Experimentally, it is possible to obtain a phase modulation of the form 
(\ref{alpha}) by simply using acousto-optical modulators and a set of radio-frequency 
generators.  The exact technical realization is of no particular interest here, 
and we refer to Ref.~\cite{schiavoni} for further details. We only notice that 
it is possible experimentally to carefully control the phase difference $\phi$ 
between the two harmonics. This allows us to carefully control the symmetry of the 
system (see Sec.~\ref{symmetric}).

\begin{figure}[ht]
\begin{center}
\mbox{\epsfxsize 3.5in \epsfbox{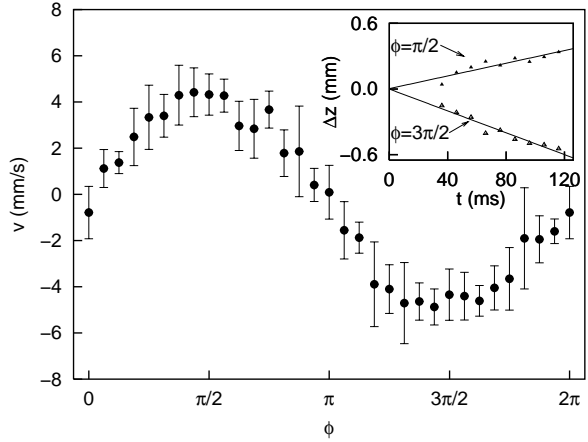}}
\end{center}
\caption{Velocity of mass of the atomic cloud as a function of the phase $\phi$. 
Inset: displacement of the center of mass of the atomic cloud as a function of 
time for two different values of the phase $\phi$. Figure taken from 
Ref.~\protect\cite{schiavoni}.}
\label{fig9}
\end{figure}

The experiment of Ref.~\cite{schiavoni} clearly demonstrated the control of the
current through a spatially symmetric potential by varying the time-symmetries
of the system. In that work, the dynamics of the atoms in the optical lattice 
was studied by direct imaging of the atomic cloud with a CCD camera. For a
given phase $\phi$ the position of the center of mass of the atomic cloud was 
studied as a function of time. It should be noticed that in principle it is necessary
to transform the measurements from the laboratory reference frame to the 
accelerated reference frame of the optical potential, by using the coordinate
transformation $z'=z-\alpha(t)/2k$. However in the case of Ref.~\cite{schiavoni} 
this is not necessary as for the typical time scales of that experiment 
(period of the ac force and imaging time) the measured positions of the c.m. of the
atomic cloud in the laboratory and in the accelerated reference frame are 
approximately equal. The results of that experiment are reported in 
Fig.~\ref{fig9}. It can be seen that the center of mass of the atomic cloud moves
with constant velocity (see inset). This velocity shows the expected dependence 
on the phase $\phi$: for $\phi=n\pi$, with $n$ integer, the velocity (current 
of atoms) is zero, while for $\phi=\pi/2, 3\pi/2$ the velocity reaches a maximum
(positive or negative). This because although the symmetry $F(t+T/2)=-F(t)$ is
broken for any value of the phase $\phi$, there is a residual symmetry 
$F(t)=F(-t)$ which forbids the current generation. This symmetry is controlled
by the phase $\phi$: for $\phi=n\pi$ it is realized, while for $\phi=(2n+1)\pi/2$ 
it is maximally broken. 
\\
\\
The experiment of Ref.~\cite{schiavoni} constitutes therefore a clear demonstration 
of the ideas of Sec.~\ref{symmetric}: the atoms can be set into directed motion 
through a symmetric potential by breaking the temporal symmetry of the system.
We notice that the described experiment reproduces well the dependence of the 
current on the phase $\phi$ derived in Sec.~\ref{symmetric} on the basis of the
analysis of symmetries which apply in the Hamiltonian limit, i.e. in the 
absence of dissipation. This because the experiment of Ref.~\cite{schiavoni}
was made in the regime of relatively strong driving, which well approximates the 
Hamiltonian regime. The role of the dissipation in a spatially symmetric system
was examined in a subsequent investigation \cite{jones}. In that work the 
possibility to rectify fluctuations in a spatially symmetric optical lattice was 
also demonstrated, i.e. a Brownian motor was realized.

In order to discuss the role of fluctuations, it is essential first to introduce 
the quantity appropriate to describe the amplitude of the fluctuations in an 
optical lattice. We have previously seen that in an optical lattice fluctuations 
are determined by optical pumping between different ground state sublevels. 
Therefore the appropriate quantity to describe the amplitude of the fluctuations
is the {\it optical pumping rate}, i.e. the number of optical pumping cycles per
unit time.

In the work of Ref.~\cite{jones} the current of atoms through the lattice was
studied as a function of the applied ac force for different values of the 
optical pumping rate $\Gamma'$. It should be notice that in an optical lattice
the optical pumping rate can be varied while keeping constant the optical 
potential. Indeed for an intensity $I$ of the lattice beams, and a detuning
$\Delta$ from atomic resonance, the optical pumping rate is proportional to
$\Gamma '\propto I/\Delta^2$ while the depth $U_0$ of the optical potential scales
as $U_0\propto I/\Delta$. Therefore by changing simultaneously $I$ and $\Delta$ 
it is possible to keep $I/\Delta$ constant while varying $I/\Delta^2$, i.e.
it is possible to change the optical pumping rate while keeping constant the 
optical potential. Experimentally it is possible to verify that the potential
is constant by using pump-probe spectroscopy \cite{yanko}. This technique indeed
allows a precise measurement of the oscillation frequency at the bottom of the
potential wells.

\begin{figure}[ht]
\begin{center}
\mbox{\epsfxsize 3.5in \epsfbox{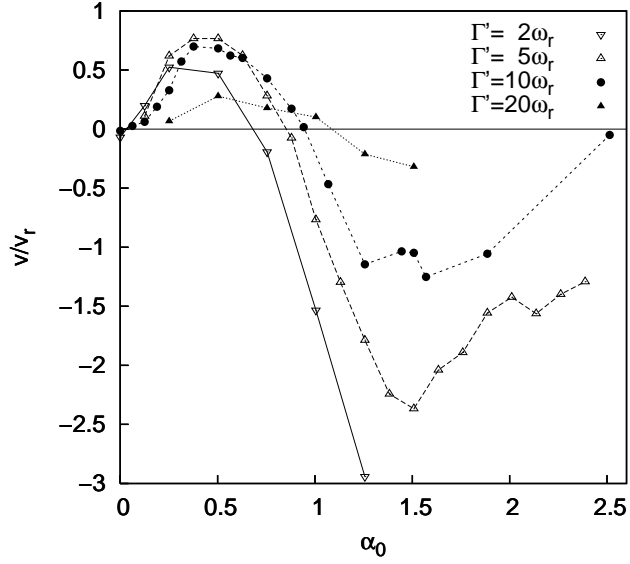}}
\end{center}
\caption{Numerical simulations for a sample of atoms in a 1D lin$\perp$lin
optical lattice. The average atomic velocity, in units of the recoil velocity
$v_r=\hbar k/m$, is shown as a function of the amplitude of the phase modulation.
Different date sets correspond to different optical pumping rates $\Gamma'$,
expressed in terms of the recoil frequency $\omega_r=\hbar k^2/2m$.
The optical potential is the same for all data. The relative phase between
the two harmonics is $\phi=\pi/2$, a choice which breaks the time symmetry 
of the system.}
\label{fig10}
\end{figure}

\begin{figure}[ht]
\begin{center}
\mbox{\epsfxsize 3.5in \epsfbox{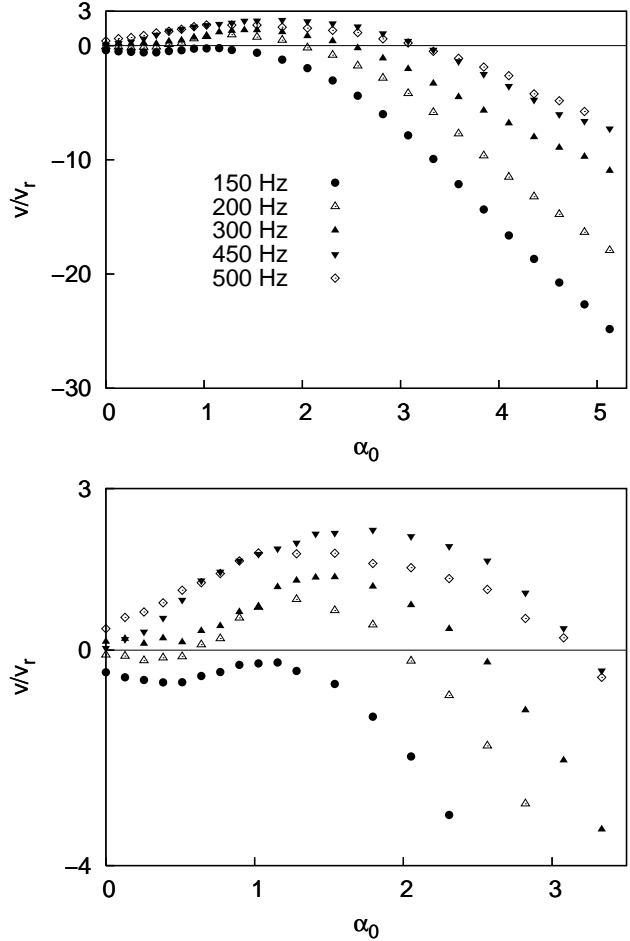}}
\end{center}
\caption{Experimental results for the atomic velocity as a function of the 
amplitude of the phase modulation. The top graph include all experimental
results, while the bottom graph evidences the region of small ac forces.
The optical potential is the same for all measurements. Different data
sets correspond to different optical pumping rate, and they are labeled
by $\Gamma_s=[\omega_v/(2\pi)^2]/\Delta$ ($\omega_v$ is the vibrational
frequency) which is proportional to the optical pumping rate.}
\label{fig11}
\end{figure}

Figures \ref{fig10} and \ref{fig11} (from Ref.~\cite{jones}) show the results
of numerical simulations and experiments. 

The experimental and numerical data are consistent and show a clear dependence
on the amplitude of the applied force and on the optical pumping rate. Consider
first the dependence on the ac force magnitude. For a small amplitude of the 
ac force the average atomic velocity is an increasing function of the force
amplitude., with the atoms moving in the positive direction. At larger amplitude
of the ac force the velocity decreases, and a current reversal is observed,
with the atomic cloud moving in the negative direction. This kind of behaviour,
named {\it current reversal}, is typical of rocking ratchets \cite{reimann02}.
We examine now the dependence of the current on the optical pumping rate, i.e.
on the noise level. We observe from Figs.~\ref{fig10},\ref{fig11} that the 
dependence of the current on the optical pumping rate is very different 
depending on the ac force amplitude. For large amplitude of the applied force
the magnitude of the current (in absolute value) is a decreasing function of 
the optical pumping rate. This means that in this regime the motion can be
attributed to deterministic forces and correspond to force rectification
by harmonic mixing: in a nonlinear medium the two harmonics, of frequency
$\omega$ and $2\omega$ and phase difference $\phi$, are mixed and the rectified
force produces a current $C\propto\sin\phi$, as discussed in Sec.~\ref{symmetric}.
In the considered experiment the nonlinearity of the medium is the anharmonicity
of the optical potential. In this regime of rectification of the forces the noise
does not play any constructive role in the generation of the current of atoms.
On the contrary, the noise disturbs the process of rectification of the forces, 
and indeed in the current decreases for increasing optical pumping rate. We
conclude that this regime does not correspond to the rectification of fluctuations.
A very different dependence of the current amplitude on the optical pumping 
rate is found in the regime of small amplitudes of the applied force. Indeed
in this regime the current is for small pumping rates an increasing function
of the pumping rate, and the current vanishes in the limit of vanishing optical
pumping rates. At larger pumping rates the current reaches a maximum and then
decreases again. This bell-shaped dependence of the current on the optical
pumping rate is a typical signature of a Brownian motor: in the absence of
fluctuations the current is zero, then increases until the fluctuations are
so large that the presence of the potential and of the applied fields become
irrelevant, and the current decreases again. We can therefore conclude that 
in the regime of small ac force amplitude the optical lattice provides an 
implementation of a Brownian motor.

\section{Conclusions}

In this review the basic ideas underlying Brownian motors have been outlined 
and two different implementations, the flashing ratchet and the rocking ratchet, 
described. The important role played by the symmetries in determining a current 
has been pointed out. The second part of this review discusses a recent 
implementation of the rocking ratchet using cold atoms in an optical lattice. 
Optical lattices are an ideal model system for statistical physics because of 
their tunability: both the optical potential and the noise level can be carefully 
controlled.  The described experiment shows well the rectification of fluctuations,
and therefore demonstrates the realization of a Brownian motor with cold atoms in 
an optical lattice.

The aim of the present review is also to give an idea of the important role that
optical lattices can play as model system to study phenomena of statistical physics. 
This goes beyond the modelling of Brownian motors. For example the 
phenomenon of {\it stochastic resonance} has been recently observed in a 
near-resonant optical lattice \cite{stoc1,stoc2}. Far-detuned optical lattice have
been used to investigate chaotic motion, both in the classical and quantum regime
\cite{chaos}. Finally, a new class of ratchets has been introduced which combines
"standard" ideas of thermal ratchets with chaos: these are deterministic chaotic
ratchets in which there are no thermal fluctuations and it is chaos which mimics
the role of noise \cite{jung}. It is not difficult to predict that optical lattices 
will play a major role in the experimental study of this new class of ratchets.

\begin{acknowledgments}
The author is grateful to EPSRC, UK and to the Royal Society for financial
support, and to Eric Lutz for critical reading the manuscript.
\end{acknowledgments}

\vspace{2cm}

\centerline{\bf Author's biography}
          
\vspace{1cm}                                                                                                                   
{\it Ferruccio Renzoni} studied Physics at the University of Pisa (Italy) where
he obtained his M.Sc. in 1993. He then obtained his Ph.D. from the Technische
Universit\"at Graz (Austria) in 1998 with a thesis on coherent population
trapping in atomic systems. He then spent two years in Germany, at the Institut
f\"ur Laserphysik of the University of Hamburg, and three years in France, at the
Laboratoire Kastler Brossel (Ecole Normale Sup\'erieure, Paris), where he
obtained his {\it Habilitation \`a diriger des recherches}. During the stay in
Germany and France, he became interested in optical lattices and in the possibility
to use them to generate models for statistical mechanics. Since 2003 he is at
the Department of Physics and Astronomy of the University College London,
where he leads the laser cooling group.

\end{document}